\documentclass[twocolumn,prb]{revtex4-2}
\usepackage{graphicx,psfrag,color}
\def\no{\noindent}
\def\bc{\begin{center}}
\def\ec{\end{center}}

\def\beq{\begin{equation}}
\def\eeq{\end{equation}}

\def\d{\downarrow}
\def\u{\uparrow}

\def\bj{{\bf j}}

\def\br{{\bf r}}

\def\bk{{\bf k}}

\def\bc{{\bf c}}

\begin{document}

\title{
Edge modes in chiral electron double layers 
}

\author{Klaus Ziegler$^{1}$ and Roman Ya. Kezerashvili$^{2,3}$}
\affiliation{
$^1$Institut f\"ur Physik, Universit\"at Augsburg, D-86135 Augsburg, Germany\\
$^{2}$Physics Department, New York City College of Technology, The City University of New York,
Brooklyn, NY 11201, USA\\
$^{3}$The Graduate School and University Center, The City University of New York,
New York, NY 10016, USA
}
\date{\today}

\begin{abstract}

We study the quasiparticles in chiral double layers with electron pairing
within the framework of the Bogoliubov de Gennes equation. 
In the presence of an edge it is demonstrated that the quasiparticle modes can be distinguished as edge modes
and bulk modes, which appear at different energies. The bulk-edge correspondence is obtained 
by an analytic continuation from the in-gap edge modes to the bands of bulk modes. 
By varying the energy we find a transition from localized edge modes to delocalized bulk modes. We
calculate the quasiparticle currents, discuss briefly how these currents couple to external 
currents, and predict how this can be used to control the quasiparticle modes. 
\end{abstract}

\maketitle

Quasiparticle modes of charged particles, such as electrons, create electric currents. They 
can couple to external currents and external magnetic fields. This offers an opportunity to 
control quasiparticle modes with macrosopic fields. Moreover, edge currents in a superconductor
couple to supercurrents through the Josephson effect.
This provides an additional control of quasiparticle modes. 

Double layer heterostructures of spatially separated two-dimensional 
electrons and holes have been a subject of considerable research interest for the last few decades.
Inspired by the pairing effect in electron-hole double 
layers~\cite{1976JETP...44..389L,PhysRevLett.77.1564,PhysRevLett.93.266801,
PhysRevB.86.115436,Wang2019,pnas.2205845119},
we exploit the duality between the electron-hole double layer (EHDL) and the
electron double layer (EEDL) to study superconductivity 
in the latter~\cite{2020PhRvR...2c3085S,PhysRevLett.128.157001}.
Our physical system is a chiral double layer without interlayer hopping.
The electrons in such a system are subject to pairing due to geometric 
effects~\cite{2020PhRvR...2c3085S}.
The goal is to study edge modes in a circular geometry with
a circular hole on a closed surface or at the center of an infinite double layer. 
A central question addressed in the present paper concerns the existence of
the bulk-edge correspondence for the spectra and the 
eigenfunctions~\cite{Graf2013,doi:10.1142/S0129055X20300034}.

\begin{figure}[t]
\begin{center}
\includegraphics[width=0.4\linewidth]{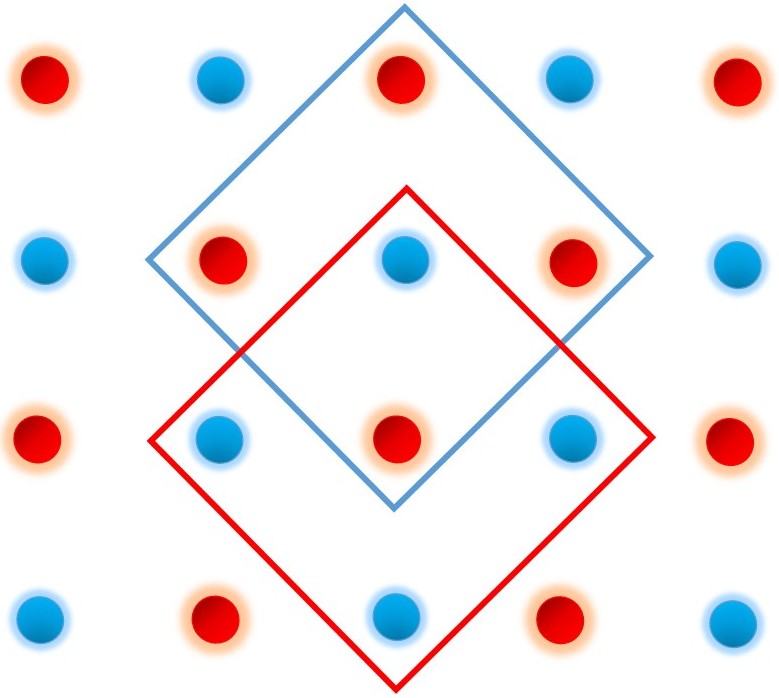}
\caption{
Top view of the EEDL.
The effective attraction between electrons in the top (red) and the bottom layer (blue) 
is caused by geometry. The blue (red) square attracts the blue (red) electron in the center 
to the four red (blue) electrons in the corners.
Here we assume that the Fermi energy is in the gap between the two bands. This implies that the filling factor is $1/2$, 
i.e., half of the lattice is occupied in each layer. 
}
\label{fig:eff_attraction}
\end{center}
\end{figure}

The EEDL comprises two spatially separated electronic layers with a positively charged extra layer, 
which acts as a gate and controls charge neutrality. 
The latter can either be an external gate or is provided by the positive charges inside the metallic layers.
In both cases the entire system preserves charge neutrality. The electrons in the two layers repel each
other due to the Coulomb interaction. A geometric constraint enables the electrons 
to form interlayer Cooper pairs. This is formally linked to the fact that the EEDL is dual to an
EHDL~\cite{2020PhRvR...2c3085S,https://doi.org/10.1002/ctpp.202300014}, where in the latter
electron-hole pairs are created by an attractive interlayer Coulomb interaction~\cite{1976JETP...44..389L}.
On the level of the Bogoliubov de Gennes Hamiltonian of the quasiparticles, the duality is a unitary mapping
described in App. \ref{app:diag}.  
This effect can be understood as a geometrically induced attractive interaction between equally 
charged particle due to the double-layer structure. The Coulomb interaction repels the 
electrons in each layer, as well as the electrons in different layers. At minimal Coulomb energy 
we get an ordered structure as illustrated in Fig. \ref{fig:eff_attraction}. The electron at the center of the
square is repelled by the four electrons in the corners of the square, 
which appears as if it is trapped. The same effect exist when we assume an attractive interaction
between the center electron and the corner electrons.

The quasiparticles of the EEDL
are described by the Bogoliubov de Gennes Hamiltonian of two layers 
with opposite chiralities~\cite{PhysRevLett.128.157001}
\begin{equation}
\label{hamiltonian0}
H^{}_{\rm EEDL}(\Delta)
=\pmatrix{
H & \Delta\sigma_2 \cr
\Delta^*\sigma_2 & -H^*\cr
}
,
\end{equation}
where $H=h_1\sigma_1+h_2\sigma_2+h_3\sigma_3$ with Pauli matrices $\sigma_j$.
$h_{1,2}$ are nearest neighbor tight-binding hopping terms and
$h_3$ is a constant with an additional next-nearest neighbor hopping term. 
This Hamiltonian acts on a space that is spanned by four-component wave functions 
$\Psi(\br)=(\psi_{\u1}(\br),\psi_{\u2}(\br),\psi_{\d1}(\br)\psi_{\d2}(\br))^T$ on 
a two-dimensional lattice. The latter has an internal structure,
indicated by $1$ and $2$, while $\u$ and $\d$ refers to the top and bottom layer, respectively.
The spatially uniform pairing order parameter $\Delta$ can be obtained self-consistently from the
interacting EEDL system, when the Fermi energy is between the two bands of $H$~\cite{1976JETP...44..389L}.
For the following we simply consider a uniform solution of the self-consistent equation.
$H_{\rm EEDL}(0)$ is invariant under $SU(2)$ transformations (cf. Eq. (\ref{EE-EH}) in 
App. \ref{app:diag}), while the $\Delta$ matrix part breaks this invariance~\cite{read93,fradkin13}.
Finally, we assume a honeycomb lattice with broken time-reversal invariance.
The latter can either be caused by a periodic magnetic flux~\cite{PhysRevLett.61.2015} or by a 
spin texture~\cite{Hill_2011} that provides a single Dirac node in the dispersion.
In this case the internal lattice structure is associated with the two equivalent triangular 
sublattices of the honeycomb lattice. 
Then we can reduce $h_{1,2}\to i\partial_{1,2}$ and $h_3\to0$ near the Dirac node
in the continuum limit.
The Hamiltonian matrix in Eq. (\ref{hamiltonian0}) can be mapped by a unitary transformation $W$
onto a block-diagonal matrix with two Dirac matrices as (cf. App. \ref{app:diag})
\beq
\label{dirac_map}
H_{\rm EEDL}(\Delta)= W^\dagger H_{\rm Dirac}(|\Delta|)W
,
\eeq
where $H_{\rm Dirac}$ is the $2\times 2$ block diagonal matrix ${\rm diag}(h_D(|\Delta|),h_D(-|\Delta|))$
with $h_D(|\Delta|)=i\partial_1\sigma_1+i\partial_2\sigma_2+|\Delta|\sigma_3$ 
and $\Delta=|\Delta|e^{i\phi}$.
The unitary matrix $W$ depends on the phase of $\Delta$ and is explained in Eq. (\ref{W-transformation}) 
of App. \ref{app:diag}:
\beq
W
=\frac{1}{\sqrt{2}}\pmatrix{
e^{-i\phi/2}\sigma_0 & e^{i\phi/2}\sigma_1\cr
-e^{-i\phi/2}\sigma_0 & e^{i\phi/2}\sigma_1\cr
}
.
\eeq
This mapping reveals that the Hamiltonian $H_{\rm EEDL}(|\Delta|)$,
which describes the coupling of the two layers by the pairing order parameter $\Delta$, is
equivalent to two Dirac Hamiltonians $h_D(\pm|\Delta|)$. 
Thus, the triality of the Hamiltonians $H_{\rm EEDL}$, $H_{\rm EHDL}$ and $H_{\rm Dirac}$

\setlength{\unitlength}{0.8cm}
\begin{picture}(6,3) 
\put(1,2){$H_{\rm EEDL}$}
\put(3,2.2){\vector(1,0){2}}
\put(5,2.2){\vector(-1,0){2}}
\put(5.5,2){$H_{\rm EHDL}$}
\put(2,1.5){\vector(1,-1){1}}
\put(3,0.5){\vector(-1,1){1}}
\put(6.2,1.5){\vector(-1,-1){1}}
\put(5.2,0.5){\vector(1,1){1}}
\put(3.5,0.25){$H_{\rm Dirac}$}
\end{picture}

\no
enables us to calculate the eigenfunctions of one Hamiltonian and transform them to any of the other
Hamiltonians. The duality of $H_{\rm EEDL}$ and $H_{\rm EHDL}$ was used previously to explain
pairing in an electronic double layer~\cite{2020PhRvR...2c3085S}, while the mapping of 
Eq. (\ref{dirac_map}) yields a connection to the massive Dirac Hamiltonian.
For the following it is convenient to determine the eigenfunctions of the Dirac Hamiltonian first,
since it separates for the two layers (cf. App. \ref{app:Dirac_eigenvector}).

Now we consider a circular geometry, assuming that each layer is either a disk or
has a circular hole but is closed otherwise, forming a sphere, a torus, 
or is an infinite plane. In these cases we have a circular edge.
Using polar coordinates $(x,y)=(r\cos\alpha,r\sin\alpha)$, we rewrite the Dirac 
Hamiltonian for one layer as
\beq
\label{dirac01}
h_D(|\Delta|)=
\pmatrix{
|\Delta| & ie^{-i\alpha}(\partial_r-\frac{i}{r}\partial_\alpha) \cr
ie^{i\alpha}(\partial_r+\frac{i}{r}\partial_\alpha) & -|\Delta| \cr
}
\eeq
and solve the eigenvalue problem $h_D(|\Delta|)\phi_E=E\phi_E$. 
The eigenfunctions for an infinite disk with a circular hole of radius $r_0$ at the center
can be expressed by modified Bessel functions $K_n(cr)$ as
\beq
\label{Phi_E0}
\phi_{E,n}(r,\alpha)=A_n\pmatrix{
K_n(cr) \cr
i\rho K_{n+1}(cr) e^{i\alpha}\cr
}
e^{in\alpha}
\eeq
with the conditions
(see Eq. (\ref{sign_relation1}) in App. \ref{app:Dirac_eigenvector})
\beq
\label{sign_relation}
c\rho=E-|\Delta|
, \ \ 
-c=\rho(E+|\Delta|)
.
\eeq
The eigenvalues are degenerate with respect to angular quantum number
$n=0,\pm 1,\ldots$, since $h_D(|\Delta|)$ is rotational invariant
and the angular momentum operator commutes with $h_D(|\Delta|)$.
Subsequently we consider a solution with a fixed $n$, which
is simultaneously an eigenfunction of $h_D(|\Delta|)$ and the angular
momentum operator. 
After the removal of the global phase factor $e^{i(n+1/2)\alpha}$,
which does not affect quadratic forms of the eigenfunctions in expectation values,
we get from the eigenfunction in Eq. (\ref{Phi_E0})
\beq
\label{Phi_E}
{\tilde\phi}_{E,n}(r,\alpha)=A_n\pmatrix{
K_n(cr)e^{-i\alpha/2} \cr
i\rho K_{n+1}(cr) e^{i\alpha/2}\cr
}
.
\eeq
This function obeys the spinor relation 
${\tilde\phi}_{E,n}(r,\alpha+2\pi)=-{\tilde\phi}_{E,n}(r,\alpha)$;
i.e., the spinor wave function reverses its sign after a full rotation.

The relation in Eq. (\ref{sign_relation}) implies $\rho^2=(|\Delta|-E)/(|\Delta|+E)$, 
the energy $E_c=\pm\sqrt{|\Delta|^2-c^2}$ and $0\le c\le |\Delta|$.
$\rho^2$ is not symmetric with respect to $E$: it diverges for $E\sim-|\Delta|$ and it vanishes at $E=|\Delta|$. 
The sign of $\rho$ is fixed by the relations (\ref{sign_relation}), namely $\rho\le 0$
due to $c\ge0$. The non-symmetric form of $\rho$ reflects the orthogonality of the wavefunctions
for $\pm E$, while $E=0$ is an exceptional point with two coalescent wavefunctions.  

\begin{figure}[t]
\begin{center}
\includegraphics[width=0.5\linewidth]{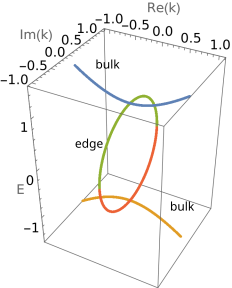}
\caption{
Bulk and edge spectra for a complex wavenumber $k$ with $Im(k)=c$.
}
\label{fig:double_layer}
\end{center}
\end{figure}

For a full disk of finite radius the solution is analogous but we must replace the modified Bessel function
$K_n$ by the modified Bessel function $I_n$ in order to get a finite normalizable
eigenfunction for $r\sim0$~\cite{abramowitz+stegun}. 
These solutions represent modes bound to the edge with energies $-|\Delta|\le E\le|\Delta|$.

Besides these edge solutions, we also obtain scattering solutions after the 
analytic continuation $c\to \pm ik$ ($k\ge 0$). The different signs do not affect the spectrum 
$E_k=\pm \sqrt{|\Delta|^2+k^2}$ but yield different eigenfunctions:
the relation~\cite{abramowitz+stegun}
\beq
\label{hankel}
K_n(-iz)
=\frac{\pi e^{i\pi (n+1)/2}}{2}[J_n(z)+iY_n(z)]
\]
\[
=e^{-in\pi/2-i\pi/4}\sqrt{2}\frac{e^{iz}}{\sqrt{\pi z}}+O\left(\frac{1}{z}\right)
,
\eeq
provides the bulk eigenfunctions in terms of the Bessel functions $J_n$ and $Y_n$.
Thus, the analytic continuation $c\to \pm ik$ yields a bulk-edge correspondence
for the eigenfunctions and the spectrum.
There is freedom to interpret $k$ as a function of physical variables. 
In our example of Dirac particles we use the wavevector $\bk$ as $k=|\bk|$.
In general, the bulk-edge correspondence can connect the same edge mode with different bulk modes,
which means that the edge modes are robust under a change of the bulk modes.
The two values $\pm ik$ are important, since it provides a superposition of two
eigenfunctions to match the boundary conditions.
The energies of the bulk solutions $E_k=\pm \sqrt{|\Delta|^2+k^2}$
represent the usual energy bands of massive Dirac particles in 2D (cf. Fig. \ref{fig:double_layer}).
The sign of $\rho$ is determined again by the relation (\ref{sign_relation}),
after the analytic continuation $c\to \pm ik$: 
\beq
\label{sign_relation2}
i\rho=\pm(E-|\Delta|)k=\pm k/(E+|\Delta|)
.
\eeq
For the upper sign the top band $E\ge|\Delta|$ has $i\rho\ge 0$, while the bottom
band $E\le -|\Delta|$ has $i\rho\le 0$. For the lower sign we get $i\rho\le 0$
for the top band and $i\rho\ge 0$ for the bottom band. This reflects the opposite
sign of the Chern numbers in the two bands~\cite{fradkin13}.

The edge modes decay exponentially on the scale $1/c=1/\sqrt{|\Delta|^2-E^2}$.
This scale diverges as $E^2$ approach the spectral boundary $|\Delta|^2$ of the edge modes. 
On the other hand, the wavevector of the bulk states vanishes as $k=\sqrt{E^2-|\Delta|^2}$
for $E^2\sim|\Delta|^2$.
Therefore, $E=\pm|\Delta|$ are singular points in the spectrum, where the edge modes become 
extended over the entire 2D system. This is reminiscent of a localization-delocalization transition
and reflects the critical energy dependence of the edge modes and its transition to bulk modes.

With the solution ${\tilde\phi}_{E,n}$ of the Dirac Hamiltonian in Eq. (\ref{Phi_E}) we 
return to the eigenvalue problem of the Hamiltonian $H_{\rm EEDL}$.
The coefficients $A_n$ and $A'_n$ of the two chiral layers must be determined through 
boundary conditions and the normalization. Assuming that the layers are similar, we choose 
$|A'_n|=|A_n|$ and $A'_n=e^{i\psi}A_n$.
Then the transformation 
in Eq. (\ref{dirac_map}) gives the eigenfunction of $H_{\rm EEDL}$ 
for the circular edge in the double layer as $e^{i(n+1/2)\alpha}\Psi_{E,n}$ with
\beq
\label{k'2}
\Psi_{E,n}=\frac{|A_n|}{\sqrt{2}}\pmatrix{
(1-e^{i\psi})K_n e^{-i\alpha/2} \cr
i(\rho-e^{i\psi}\rho^{-1})K_{n+1}e^{i\alpha/2} \cr
i(\rho+e^{i\psi}\rho^{-1})K_{n+1}e^{i\alpha/2} \cr
(1+e^{i\psi})K_n e^{-i\alpha/2}\cr
}
.
\eeq
The radial decay as well as the parameters $c$ and $\rho$ are inherited from the 
Dirac solution in Eq. (\ref{Phi_E}). Therefore, the bulk-edge correspondence
is also inherited from the Dirac solution. Moreover, the phase $\psi$
enables us to control the contributions of the Dirac solutions of the top and the
bottom layer. An example is $\psi=0$, which suppresses the zero mode, where $\rho=-1$, 
in the top layer:
\beq
\Psi_{0,n}=\sqrt{2}|A_n|\pmatrix{
0 \cr
0 \cr
-iK_{n+1}e^{i\alpha/2} \cr
K_ne^{-i\alpha/2} \cr
}
.
\eeq
Alternatively, the zero mode in the bottom layer is suppressed for $\psi=\pi$.
The actual values of the
boundary conditions are determined by the coupling of the double layer to its environment.

\begin{figure}[t]
\begin{center}
\includegraphics[width=0.75\linewidth]{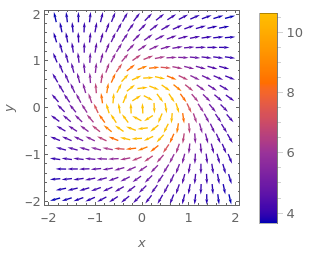}
\caption{
Bulk current density in a single chiral disk for $E>|\Delta|$ and $i\rho>0$, as follows from Eq. (\ref{sign_relation2}).
For $E<-|\Delta|$ we get the opposite vorticity. This reflects
the different Chern numbers of the top and the bottom band.
}
\label{fig:bulk_current} 
\end{center}
\end{figure}

\begin{figure}[t]
\begin{center}
\includegraphics[width=0.8\linewidth]{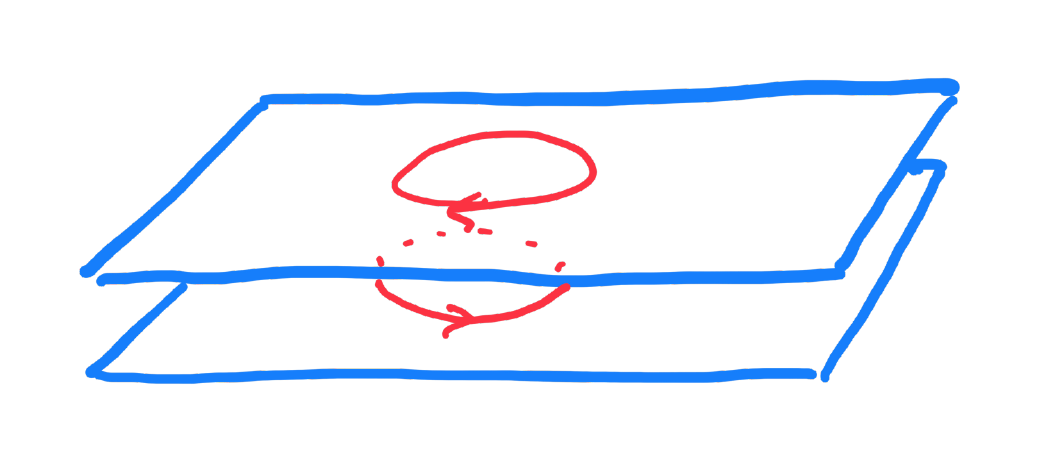}
\caption{
Chiral electronic double layer: circular edges  create counterflowing currents. 
}
\label{fig:currents}
\end{center}
\end{figure}

Charged quasiparticles, such as electrons, can be observed by the fact that they carry an electric current. 
This provides also an opportunity to create them by coupling the EEDL to an external 
current or magnetic field. 
The current operator for a single Dirac layer reads $j_\mu=(e/i\hbar)[h_{D},r_\mu]$, which gives for 
$E^2\le|\Delta|^2$ the expectation value of the edge current density
\beq
\bj_D=|A_n|^2\rho K_nK_{n+1}\pmatrix{
-\sin\alpha \cr
\cos\alpha \cr
}
,
\eeq
where $\rho<0$ implies that the current flows clockwise. According to the relation
(\ref{sign_relation}), a sign reverse $|\Delta|\to-|\Delta|$ implies a sign reverse of $\rho$, which leads to
a counter-clockwise current flow. 
The bulk current density is also represented
by a vortex field, as illustrated in Fig. \ref{fig:bulk_current},
which is a consequence of the angular momentum eigenfunction. 
The sign change of $i\rho$ when we switch from the top to the bottom band
implies a reversal of the vorticity. This reflects the opposite signs of the
Chern numbers in the two bands. Since the sign of $i\rho$ depends on
the analytic continuation $c\to ik$ or $c\to -ik$, the vorticity of the
edge currents agree either with the bulk currents of the top or the bottom
band. 

The quasiparticle current can also be calculated from the expectation 
value of the current operator 
$j_\mu=(e/i\hbar)[H_{\rm EEDL},r_\mu]$ with respect to $\Psi_{E,n}$. 
This gives for the edge current densities of the two layers
\beq
{\bf j}_{\u,\d}=C_{\u,\d}\pmatrix{
\cos(\alpha+\varphi_{\u,\d})\cr
\sin(\alpha+\varphi_{\u,\d})\cr
}
,
\eeq
where $\u$ ($\d$) refers to the top (bottom) layer.
The coefficients read
\beq
C_{\u,\d}=\pm|A_n|^2|(1\mp e^{-i\psi})(\rho\mp e^{i\psi}\rho^{-1})|K_nK_{n+1}
,
\eeq
where the different signs imply opposite current directions in the two layers.
The magnitude of the currents depends on the phase $\psi$.  
Moreover, the factor $K_n(cr)K_{n+1}(cr)$ is responsible for the radial decay 
of the current density. Finally, there is a phase shift:
\beq
\varphi_\u=\pi/2+{\rm arg}[(1-e^{-i\psi})(\rho-e^{i\psi}\rho^{-1})]
,
\]
\[
\varphi_\d=\pi/2+{\rm arg}[(1+e^{i\psi})(\rho+e^{-i\psi}\rho^{-1})]
.
\eeq
In the special case of $E=0$ we have $\rho=-1$, $\varphi_\u=\varphi_\d=\pi/2$ 
and $C_{\u,\d}=\pm 2|A_n|^2(1\mp\cos\psi)K_nK_{n+1}$.
Due to the different signs of $C_{\u,\d}$ the currents flow clockwise 
in the top layer and anti-clockwise in the bottom layer around the hole (cf. Fig. \ref{fig:currents}). 
These expressions indicate that the phase $\psi$ of the boundary conditions
can be used to control the edge currents.

The quasiparticle edge current density is balanced with the supercurrent density further away from 
edge, based on a continuity 
equation~\cite{PhysRevB.25.4515,SPUNTARELLI2010111,PhysRevB.106.L220503,PhysRevLett.128.157001},
while the quasiparticle currents of the bulk eigenfunctions are independent of the supercurrents. 
Moreover, a macroscopic edge current radiates energy from the system to the environment.
Therefore, quasiparticles would either disappear after some time, leading to $A'_n=A_n=0$
as an equilibrium state, or they are balanced by an external supply of energy, for instance, by 
a magnetic flux.

In conclusion, we have studied quasiparticle edge modes in a superconducting EEDL.
By exploiting the duality of the Hamiltonians $H_{\rm EEDL}$ and $H_{\rm Dirac}$ the
simultaneous eigenfunctions of $H_{\rm EEDL}$ and the angular momentum operator were calculated.
It turned out that the edge modes in the layers can be controlled separately, they can be 
switched off individually in the two layers by changing the boundary conditions at the edge.
The edge modes carry a current density, which can couple to external fields.
This predicts an experimental control of the quasiparticle edge modes. 
Moreover,
an advantage of the superconducting EEDL is that the quasiparticle currents are balanced
with the supercurrents as a result of charge conservation. In other words,
the edge currents are maintained by the superconducting currents. Finally, there is a 
bulk-edge correspondence, which is provided by an analytic continuation.
This correspondence characterizes two critical energies $E=\pm|\Delta|$, where a transition 
from edge modes to bulk modes and vice versa appear. These transitions are reminiscent
of localization-delocalization transitions for the quasiparticle modes in disordered
systems. An application of these ideas to a non-uniform pairing order parameter $\Delta$ 
and to its non-Abelian fluctuations is left as a project for the future. This could be of interest 
in the study of the Josephson effect in EEDL systems. Finally, it should be noted that our approach 
can be extended to deformed edges and other variations of the model in terms of perturbation
theory, as long as the energy of the perturbation is sufficiently smaller than the gap. 
{\it Acknowledgment.}
We are grateful for inspiring discussions with Andrii Iurov in the early stage of this work.

{\it Data availability.} No data were created or analyzed in this study.


\appendix

\section{Mapping of the $H_{\rm EEDL}$ Hamiltonian}
\label{app:diag}

First, we consider the mapping from the EEDL to the EHDL
$H_{\rm EEDL}\to UH_{\rm EHDL}U^\dagger=H_{\rm EHDL}$ as
\beq
\label{EE-EH}
\pmatrix{
h_1\sigma_1+h_2\sigma_2 & \Delta\sigma_2 \cr
\Delta^*\sigma_2 & h_1\sigma_1-h_2\sigma_2\cr
}
\]
\[
\to
U\pmatrix{
h_1\sigma_1+h_2\sigma_2 & \Delta\sigma_2 \cr
\Delta^*\sigma_2 & h_1\sigma_1-h_2\sigma_2\cr
}U^\dagger
\]
\[
=\pmatrix{
h_1\sigma_1+h_2\sigma_2 & \Delta\sigma_3 \cr
\Delta^*\sigma_3 & h_1\sigma_1+h_2\sigma_2\cr
}=H_{\rm EHDL}
\eeq
with
\beq
U=\pmatrix{
\sigma_0 & 0 \cr
0 & -i\sigma_1 \cr
}
.
\eeq
We note that $H_{\rm EHDL}$ is symmetric under $SU(2)$ transformations if $\Delta=0$.

Second, we have the mapping $H_{\rm EEDL}\to H_D=WH_{\rm EEDL}W^\dagger$ as
\beq
\label{W-transformation}
\pmatrix{
h_1\sigma_1+h_2\sigma_2 & \Delta\sigma_2 \cr
\Delta^*\sigma_2 & h_1\sigma_1-h_2\sigma_2\cr
}
\]
\[
\to
W\pmatrix{
h_1\sigma_1+h_2\sigma_2 & \Delta\sigma_2 \cr
\Delta^*\sigma_2 & h_1\sigma_1-h_2\sigma_2\cr
}W^\dagger
\]
\[
=\pmatrix{
h_1\sigma_1+h_2\sigma_2 +|\Delta|\sigma_3 & 0 \cr
0 & h_1\sigma_1+h_2\sigma_2-|\Delta|\sigma_3 \cr
}
\]
with $\Delta=|\Delta|e^{i\phi}$ and
\[
W 
=\frac{1}{\sqrt{2}}\pmatrix{
e^{-i\phi/2}\sigma_0 & e^{i\phi/2}\sigma_1\cr
-e^{-i\phi/2}\sigma_0 & e^{i\phi/2}\sigma_1\cr
}
.
\eeq
This implies the mapping of the eigenvector $\Phi_E$ of $H_{\rm Dirac}$ to the eigenvector $\Psi_E$
of $H_{\rm EEDL}$ as
\beq
\label{dirac_map2}
\Psi_E=W^\dagger\Phi_E
.
\eeq

\section{Eigenfunctions of the Dirac Hamiltonian}
\label{app:Dirac_eigenvector}

With the ansatz
\beq
\label{dirac_ef}
\Phi_{E,n}(r,\alpha)=\pmatrix{
f_n(cr) \cr
g_n(cr) e^{i\alpha}\cr
}
e^{in\alpha}
\eeq
we can write for the eigenvalue problem $h_D\Phi_{E,n}=E\Phi_{E,n}$ the equation
\beq
\pmatrix{
i(cg_n\rq{}+\frac{n+1}{r}g_n) \cr
i(cf_n\rq{}-\frac{n}{r}f_n) \cr
}
=\pmatrix{
(E-|\Delta|)f_n\cr
(E+|\Delta|)g_n \cr
}
.
\eeq
With the recurrence relations of the modified Bessel functions~\cite{abramowitz+stegun}
\beq
K_0\rq{}=-K_1
\ ,\ \
K_1\rq{}=-K_0-\frac{1}{cr}K_1
\]
\[
K_n\rq{}=-K_{n-1}-\frac{n}{cr}K_n=-K_{n+1}+\frac{n}{cr}K_n
\eeq
we get for $f_n=K_n$ and $g_n=i\rho K_{n+1}$ the relations
\beq
\pmatrix{
ic(g_n\rq{}+\frac{n+1}{cr}g_n) \cr
ic(f_n\rq{}-\frac{n}{cr}f_n) \cr
}
=\pmatrix{
-c\rho(-K_n) \cr
ic(-K_{n+1}) \cr
}
\]
\[
=\pmatrix{
(E-|\Delta|)K_n\cr
(E+|\Delta|)i\rho K_{n+1} \cr
}
,
\eeq
where the second equation implies
\beq
\label{sign_relation1}
c\rho=E-|\Delta|
, \ \ 
-c=\rho(E+|\Delta|)
.
\eeq
Thus, we have $c^2=|\Delta|^2-E^2$ and $\rho^2=(|\Delta|-E)/(|\Delta|+E)$, independent of $n$.
Combining the eigenfunctions of the Dirac Hamiltonians $h_D(\pm|\Delta|)$,
we get as the eigenfunction of $H_{\rm Dirac}$
\[
\Phi_{E,n}
=\pmatrix{
AK_n(cr) \cr
A i\rho K_{n+1}(cr) e^{i\alpha}\cr
A\rq{}K_n(cr) \cr
A\rq{}i\rho^{-1} K_{n+1}(cr) e^{i\alpha}\cr
}
e^{in\alpha}
\]
and with Eq. (\ref{dirac_map2}) the eigenfunction of $H_{\rm EEDL}$ as
\beq
\Psi_{E,n}
=\frac{1}{\sqrt{2}}\pmatrix{
(A_n-A'_n)K_n \cr
i(A_n\rho-A'_n\rho^{-1})e^{i\alpha}K_{n+1} \cr
i(A_n\rho+A'_n\rho^{-1})e^{i\alpha}K_{n+1} \cr
(A_n+A'_n)K_n \cr
}e^{in\alpha}
.
\eeq

\end{document}